\documentclass[5p]{elsarticle}

\usepackage{lineno,hyperref}
\modulolinenumbers[5]

\journal{Computational Materials Science}
\usepackage{epsfig}
\usepackage{amsmath}
\usepackage[utf8]{inputenc}
\usepackage{amssymb}
\usepackage{tcolorbox}
\usepackage{booktabs}

\usepackage[english]{babel}

\newcommand{\cpp}{{C\nolinebreak[4]\hspace{-.05em}\raisebox{.4ex}{\tiny\bf ++}}}

\begin{document}

\begin{frontmatter}

\title{A Sublattice Phase-Field Model for Direct CALPHAD Database Coupling\tnoteref{t1}}
\tnotetext[t1]{\copyright 2020. This manuscript version is made available under the \href{http://creativecommons.org/licenses/by-nc-nd/4.0/}{CC-BY-NC-ND 4.0 license}}

\author[inl1]{D.~Schwen\corref{correspondingauthor}}
\ead{daniel.schwen@inl.gov}
\cortext[correspondingauthor]{Corresponding author}

\author[inl1]{C.~Jiang}
\author[inl1]{L. K.~Aagesen}

\address[inl1]{Computational Mechanics and Materials Department, Idaho National Laboratory, Idaho Falls, ID 83415, United States}

\begin{abstract}
The phase-field method has been established as a \emph{de facto} standard for
simulating the microstructural evolution of materials. In quantitative modeling the
assessment and compilation of thermodynamic/kinetic data is largely dominated by
the CALPHAD approach, which has produced a large set of experimentally and
computationally generated Gibbs free energy and atomic mobility data in a
standardized format: the thermodynamic database (TDB) file format. Harnessing
this data for the purpose of phase-field modeling is an ongoing effort encompassing a
wide variety of approaches. In this paper, we aim to directly link CALPHAD
data to the phase-field method, without intermediate fitting or interpolation steps.
We introduce a model based on the Kim-Kim-Suzuki (KKS) approach. This
model includes sublattice site fractions and can directly utilize data from TDB files. Using this approach, we demonstrate
the model on the U-Zr and Mo-Ni-Re systems.
\end{abstract}

\begin{keyword}
 phase-field \sep CALPHAD \sep automatic differentiation
  \PACS 46.15.-x \sep 05.10.-a \sep 02.70.Dh
  \MSC[2010] 65-04 \sep 65Z05
\end{keyword}

\end{frontmatter}


\section{Introduction}

In the field of mesoscale materials modeling, the phase-field method has emerged
as a well-established approach for simulating the coevolution of microstructure
and properties \cite{chen_phase-field_2002, moelans2008introduction}. Describing
the phase state and concentrations via field variables with finite-width smooth
interfaces has proven an extremely flexible approach, resulting in a broad range
of applications ranging from solidification~\cite{warren1995prediction,
karma1996phase, kim_phase-field_1999} to phase
transformation~\cite{wheeler1992phase} to grain growth~\cite{fan1997diffusion,
moelans2008quantitative}.

Quantitative phase-field modeling of realistic material systems requires
thermodynamic and kinetic input data in the form of Gibbs free energies and
atomic mobilities. The assessment and compilation of such data through a
combination of theoretical and experimental data are formalized by the CALPHAD
approach \cite{lukas2007computational}. In CALPHAD, Gibbs free energies are
expressed as phenomenological function expansions combined with semi-empirical
entropy models. As a standard machine-readable delivery format for these free
energies, the thermodynamic database (TDB) ASCII-based file format was
established. A large swath of open thermodynamic and kinetic data exists on the
web and can be explored using search engines such as TDBDB
\cite{VANDEWALLE2018173}.

Various indirect approaches exist to make this CALPHAD data available for
phase-field modeling. Offline approaches utilize external thermodynamic software
to precalculate the internal equilibration of site fractions. These
precalculated free energies can be fitted to simple parabolas
\cite{CHOUDHURY2015287} or tabulated and interpolated over the entire state
space. Tabulation can be performed on demand or on the fly to incrementally
build the data, and tabulation approaches using polyadic tensor decomposition
expansions \cite{Coutinho2020} have been developed to address the explosion of
the state space volume with increasing dimensionality. Zhang et al.
\cite{ZHANG2015156} presented a direct coupling approach to multi-sublattice
models, using an iterative two-step process to evolve phase-field variables and
site fractions. A direct one-to-one relation between the variables in the
CALPHAD database and those in the phase-field model is not established for
models other than Type I $(A,B)_k(A)_l$.

The MOOSE framework \cite{moose-web-page, permann2019moose} contains
functionality developed by the authors to extract the functional form of a free
energy from TDB files. Direct usage of CALPHAD free energies in MOOSE-based
phase-field simulations has so far been limited to models defined on a single
sublattice (i.e., the substitutional solution model). The compound energy
formalism \cite{hillert1970, Sundman1981297, HILLERT2001161} used in CALPHAD
databases permits the description of phases with multiple sublattices, each with
their  own independent concentration degrees of freedom. Such a description is
necessary to properly describe structurally complex intermetallic compounds such
as the sigma phase in the Fe-Cr system \cite{JACOB201816}. Current phase-field
models implemented in MOOSE, however, deal only in the overall composition of
phases. The basic assumption for every material point is a local thermodynamic
equilibrium.

Thermodynamic modeling software that uses multi-sublattice free energies must
therefore perform a minimization of the internal degrees of freedom under the
constraint of a given total overall concentration. The aim of this work is to
derive a phase-field model that evolves the internal degrees of freedom in
phases with multiple sublattices on-the-fly as part of a coupled partial
differential equation system. As such, no preprocessing, tabulation, fitting, or
approximation of the free energy density of the system must be performed. This method
can therefore be used as a benchmark to quantify the errors inherent in
preprocessing methods.

\section{Sublattice KKS model}

We recall that the original Kim-Kim-Suzuki (KKS) phase-field model
\cite{kim_phase-field_1999} introduces the concept of phase concentrations
$c_{ij}$ for every component $i$ and phase $j$ in the system. The phase
concentrations are assumed to be in local thermodynamic equilibrium for each pair of phases $j$ and $j'$:
\begin{equation}
  \frac{\partial f_j}{\partial c_{ij}} =\frac{\partial f_{j'}}{\partial c_{ij'}}
\end{equation}
where $f_j$ is a phase free energy density. The
physical concentration $c_i$ for component $i$ is defined as:
\begin{equation}
  c_i = \sum_j h_j c_{ij}
\end{equation}
where $h_j$ represents switching functions that may depend on any combination of
non-conserved order parameters $\eta_j$ and satisfies $\sum_j h_j = 1$.

In a phase with multiple sublattices $k$, the phase concentrations $c_{ij}$  are
split up into sublattice concentrations $c_{ijk}$, as per:
\begin{equation}
c_{ij} = \sum_k a_{jk} c_{ijk} \label{eq:phaseconc}
\end{equation}
where $a_{jk}$ is a stoichiometric coefficient denoting the fraction of $k$
sublattice sites in phase $j$. The physical concentration for component $i$ is
then written as:
\begin{equation}
c_i = \sum_j h_j \sum_k a_{jk} c_{ijk} \label{eq:physconc}
\end{equation}

The sublattice concentrations enter the Allen-Cahn and Cahn-Hilliard equations only through their sum over all sublattcices in a given phase, the phase concentration. Thus the minimization of the total free energy requires each phase to have the minimum energy partitioning of the phase concentration onto its sublattices. That means the sublattice concentrations within a phase are given by a
constrained minimization of the phase free energy density $f_j$. We use the Lagrange
multiplier technique with the following constraint:
\begin{equation}
  g_{ij}(\vec{c_{ij}}) = \left[\sum_k a_{jk} c_{ijk}\right] - c_{ij} \label{eq:constraint}
\end{equation}
where $\vec{c_{ij}}$ is the vector of $c_{ijk}$ for all $k$ and fixed $i$ and $j$. With the Lagrange
multiplier $\lambda_{ij}$, we can write
\begin{equation}
\nabla_{\vec{c_{ij}}} f_j(\vec{c_{ij}}) = \lambda_{ij}\nabla_{\vec{c_{ij}}} g_{ij}(\vec{c_{ij}})
\end{equation}
where $\nabla_{\vec{c_{ij}}}$ is the differential operator of partial derivatives for the $\vec{c_{ij}}$ directions. Taking the component wise equality of the gradient vectors we obtain:
\begin{equation}\label{eq:lambda}
\frac 1{a_{jk}} \frac{\partial f_j}{\partial c_{ijk}} = \lambda_{ij}\quad \forall k
\end{equation}

Here, we pulled the stoichiometric coefficient $a_{jk}$ over to the left-hand
side. This means that, for all sublattice pairs $k$ and $k'$:
\begin{equation}
\frac 1{a_{jk}} \frac{\partial f_j}{\partial c_{ijk}} = \frac 1{a_{jk'}} \frac{\partial f_j}{\partial c_{ijk'}} \label{eq:sublatticechempot}
\end{equation}

Taking the derivative of Eq. \ref{eq:phaseconc} with respect to the phase
concentration $c_{ij}$ yields:
\begin{equation}\label{eq:dphaseconc}
1 = \sum_k a_{jk} \frac{\partial c_{ijk}}{\partial c_{ij}}
\end{equation}

To obtain $\mu_{ij}$, the chemical potential of a constituent $i$ in phase $j$, we
take the derivative of the phase free energy density $f_j$, with respect to the phase
concentration $c_{ij}$:
\begin{equation}
\mu_{ij}  = \frac{\partial f_j}{\partial c_{ij}} = \sum_k \frac{\partial f_j}{\partial c_{ijk}} \frac{\partial c_{ijk}}{\partial c_{ij}}
\end{equation}

We can then substitute in eqs. \ref{eq:lambda} and \ref{eq:dphaseconc} to obtain
\begin{align}
  \mu_{ij} & = \sum_k \frac1{a_{jk}}\frac{\partial f_j}{\partial c_{ijk}} a_{jk}\frac{\partial c_{ijk}}{\partial c_{ij}}
  = \lambda_{ij} \sum_k a_{jk}\frac{\partial c_{ijk}}{\partial c_{ij}} \\
  & = \frac 1{a_{jk}} \frac{\partial f_j}{\partial c_{ijk}}
\end{align}
where $k$ can be an arbitrary sublattice of phase $j$, as per
Eq. \ref{eq:sublatticechempot}. We note that, for phases containing only one
sublattice, this model reduces to the original KKS model.

Note that at no point in the derivation do we require the stoichiometric
coefficients $a_{jk}$ to sum up to one ($1=\sum_k a_{jk}$). This allows us to
omit sublattices for constituents in certain phases if the constituent is not
found in a given sublattice, as according to the CALPHAD model.

\subsection{Evolution equations for an example three-phase system}
\label{sect:evolution_equations}
To demonstrate the capability of the sublattice KKS phase-field model (SLKKS),
we modified the three-phase KKS model originally described in \cite{Ohno_2010}
to include multiple sublattice concentrations. (However, it should be noted that
the sublattice formulation is not restricted in any way to three-phase systems.)
In this model, the three phases are represented by three order parameters
($\eta_1$, $\eta_2$, and $\eta_3$) constrained such that $\eta_1 + \eta_2 +
\eta_3 = 1$. The total free energy density of the system $\mathcal{F}$ is given by:
\begin{equation}
\mathcal{F} = \int_\Omega \left[ f_\text{loc} + f_\text{gr}  \right] \text{d}V
\label{eq:energy}
\end{equation}
where the local energy density $f_\text{loc}$ is given by:
\begin{equation}
f_\text{loc} = \sum_{j=1}^3  h_j f_j + W \eta_j^2 (1-\eta_j)^2
\label{eq:floc}
\end{equation}
and $W$ is the potential barrier height. The form of the switching function used for the three-phase system is described in Section \ref{subsect:switching_function}. The gradient energy density $f_\text{gr}$ is given by:
\begin{equation}
f_\text{gr} = \sum_{j=1}^3 \frac{\kappa}{2} \left| \nabla \eta_j \right|^2
\label{eq:fgrad}
\end{equation}
To enforce the constraint $\eta_1 + \eta_2 + \eta_3 = 1$, a Lagrangian $\mathcal{F}_L$ is constructed based on Eq.~\ref{eq:energy}:
\begin{equation}
\mathcal{F}_L = \int_\Omega \left[ f_\text{loc} + f_\text{gr} + \lambda \left(1-\sum_{j=1}^3 \eta_i \right) \right] \text{d}V
\label{eq:lagrangian}
\end{equation}
The Allen-Cahn equation for the evolution of each of the three phases is derived from the variational derivative of the Lagrangian:
\begin{equation}
\frac{\partial \eta_i}{\partial t} = -L \frac{\delta \mathcal{F}_L}{\delta \eta_i}
\label{eq:AllenCahn}
\end{equation}
The full form of Eq.~\ref{eq:AllenCahn} is given in Ref.~\cite{Schwen2017}.


In \cite{gyoon_kim_phase-field_2004}, a further simplification of the
Cahn-Hilliard equation in the KKS model is introduced, reducing the order of the
partial differential equation to a modified diffusion problem. In MOOSE, this
modification significantly improves the convergence of KKS multiphase
simulations. The derivation by Kim et al. can be applied in straightforward
fashion to the evolution equation, including the sublattice concentration,
yielding:
\begin{equation}
   \frac{\partial c_i}{\partial t} = \nabla\cdot D\sum_j h_j\sum_k a_{jk}\nabla c_{ijk} \label{eq:chdiff}
\end{equation}

We implemented Eqs. \ref{eq:phaseconc}, \ref{eq:sublatticechempot}, \ref{eq:AllenCahn}, and
\ref{eq:chdiff} in the \cpp Marmot \cite{tonks_object-oriented_2012}
application, which is based on the MOOSE finite element framework
\cite{Gaston_2015}.

\subsection{Switching function}
\label{subsect:switching_function}

Crucial for the construction of a multiphase model is the choice of switching
function $h_j$, which represents the physical phase fraction of a given phase
$j$. A thermodynamically consistent switching function should not introduce
artificial driving forces. In particular, the first derivative $\frac{\partial
h_j}{\partial\eta_{j'}}$ should be zero for any $j'$ if $\eta_{j'}=0$, with the
second derivative being either positive or zero. This implies the absence of a
driving force not strictly collinear with the edges of the Gibbs simplex
defining the configurational phase space. Such a switching function prevents
artificial formation of third phases along the interfaces of any two phases. We
note that physically meaningful driving forces for the nucleation of new
phases---resulting from the interplay of bulk free energy density and interfacial free
energy---are unaffected by this choice.

A switching function that satisfies the aforementioned conditions for a three-phase
system is the so-called ``tilting function'' defined by Folch and Plapp
\cite{Folch2005} as:
\begin{equation}\label{eq:tilt}
\begin{aligned}
  h'_j(\eta_j, \eta_{j'}, \eta_{j''}) = \frac{\eta_j^2}{4} \left( 15(1 - \eta_j) [1 + \eta_j - (\eta_{j''} - \eta_{j'})^2] \right. \\
                                                                  \left. + \eta_j(9\eta_j^2 - 5) \right)
\end{aligned}
\end{equation}
where $\eta_j, \eta_{j'}, \eta_{j''}$ represents the cyclic rotations of the set of three
order parameters associated with the three phases of the system.

We observe that numerical instability can arise from the formulation in
\ref{eq:tilt}, as $h'_j$ can become either negative or larger than one for
certain combinations of $\eta_j, \eta_{j'}, \eta_{j''}$, resulting in unphysical
phase fractions and divergent negative order parameters corresponding to phases
with large free energies (such as line compound phases in composition space
regions away from the compound stoichiometry). $h'_j$ values outside of the
interval $[0,1]$ require at least one order parameter to assume values outside
the interval $[0,1]$. If  an unconstrained partial differential equation (PDE)
solver is used to evaluate the time evolution of the phase-field equations, the
absence of physical barriers can lead the solve into these unphysical regions of
the phase space.

To mitigate this problem, we propose a small modification of the tilting
functions in order to effectively constrain the range of the function's arguments by
passing them through a soft Heaviside function $h''$ defined as:
\begin{equation}
h''(\eta)=
  \begin{cases}
    0 , & \eta \leq 0 \\
    3\eta^2-2\eta^3 , & 0 < \eta < 1 \\
    1 , &1 \leq \eta
  \end{cases}
\end{equation}
Thus, we define $h_j = h'_j\left(h''(\eta_j), h''(\eta_{j'}),
h''(\eta_{j''})\right)$, and, through application of the chain rule, it follows
trivially that this formulation satisfies the condition of the absence of an
artificial driving force.

\section{Example applications}

We use the Python package pycalphad \cite{otis_pycalphad:_2017} to read and
parse TDB files. Pycalphad utilizes the SymPy \cite{10.7717/peerj-cs.103}
symbolic algebra Python package to construct abstract syntax trees (ASTs) of the
free-energy expressions for each phase found in the TDB files. An AST is a tree
data structure representing a mathematical expression, including all variables,
operators, and functions used therein. We implemented a SymPy \emph{printer} to
output the free energy density expression AST in a text format compatible with the
function expression parser \cite{fparser-web-page} used in the MOOSE framework.

The resulting function expressions for the phase free energies as a function of
their respective sublattice concentrations can be pasted directly into MOOSE
input files. As laid out in \cite{Schwen2017}, the expressions are then parsed
at run time, and symbolic automatic differentiation is performed to obtain the
expressions for the chemical potentials and derivatives needed to construct the
Jacobian matrix of the problem. The free energy density expression and its derivatives
are then transformed into executable machine code through just-in-time
compilation for high-performance evaluation.

\subsection{Uranium-zirconium}

We demonstrate the sublattice KKS model on the binary uranium-zirconium system
as assessed by Quaini et al. \cite{QUAINI2018104}. The TDB file for the
system was obtained from the Thermodynamics of Advanced Fuel International
Database \cite{taf-id-web-page}.

At a temperature of 750 K, only the orthorhombic $\alpha$-uranium, hcp
$\alpha$-zirconium, and $\delta$-UZr phases are stable. The former two phases
are modeled using single sublattice models, while the $\delta$-UZr phase is
described using the following sublattice model:
$(U,Zr)_{k_{\delta,0}}(U,Zr)_{k_{\delta,1}}$ with $k_{\delta,0} = \frac13$ and
$k_{\delta,1} = \frac23$.

\begin{figure}[tbp]
\centering
  \epsfig{width=\linewidth,file=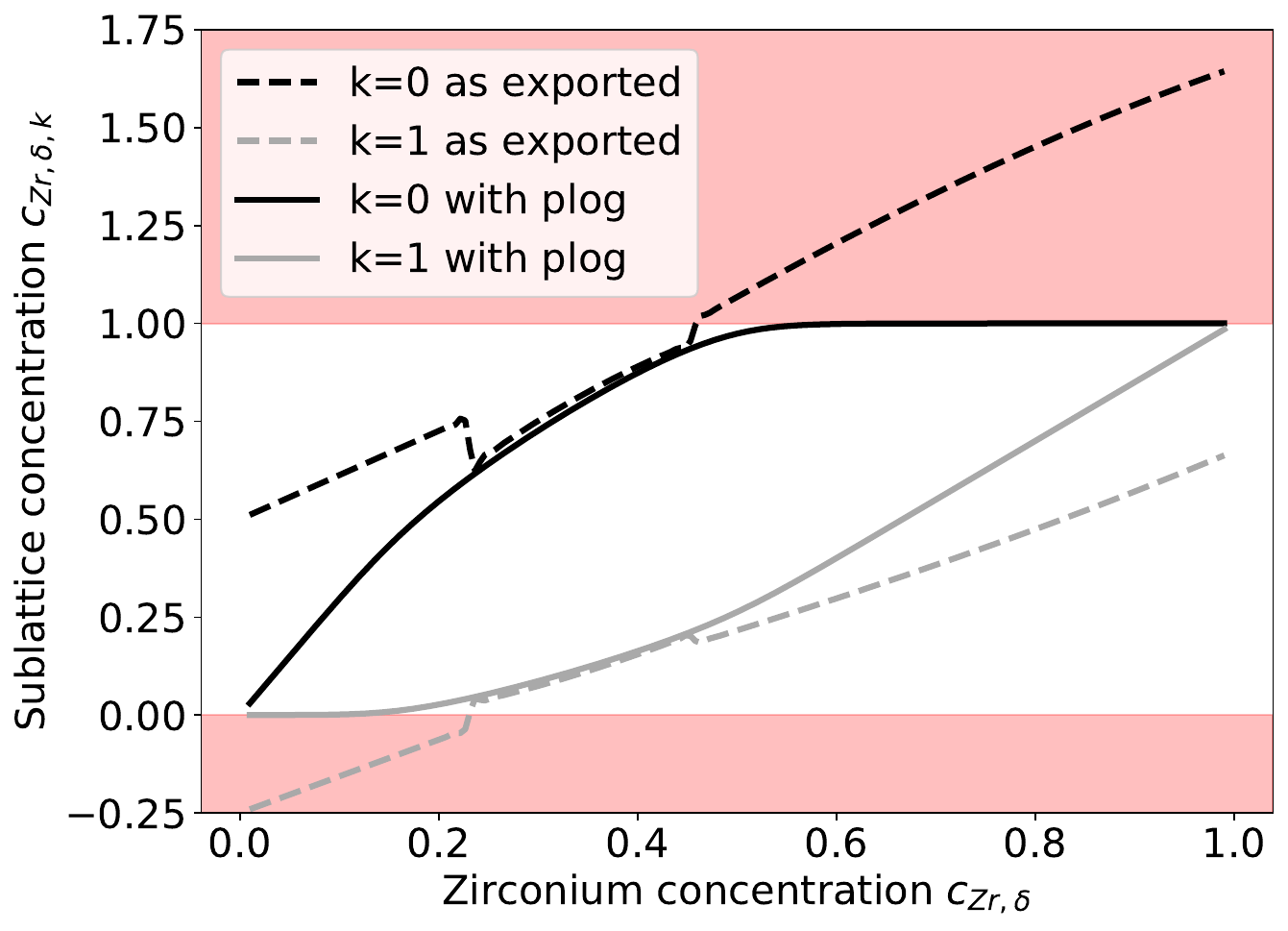}
  \caption{\label{fig:sublattice_concentrations} Sublattice concentrations in the uranium-zirconium $\delta$-phase as a function of phase concentration $c_{Zr,\delta}$. Shaded areas denote unphysical concentration regimes.}
\end{figure}

When assembling the ideal mixing free energy density part of the AST, pycalphad inserts
terms containing the product of site fraction $y_i$ and its logarithm as:
\begin{equation}
  \begin{cases}
      y_i\log y_i & y_i >  \epsilon_y \\
      0 & y_i \leq \epsilon_y
   \end{cases}
   \label{eq:pycalphad_ideal}
\end{equation}
where $\epsilon_y$ is a minimum site fraction constant defined as $10^{-16}$.
While this procedure improves the numerical stability in pycalphad, it is
detrimental in the finite-element-based implicit MOOSE solves. If during the
solve, a site fraction value slips below $\epsilon_y$, the local chemical
potential and its gradient switch to zero and do not contribute to a driving
force that takes the site fraction back to physical values. This is illustrated in
Fig. \ref{fig:sublattice_concentrations} (dashed curves), where a solve for
the sublattice site fractions in the uranium-zirconium $\delta$-phase as a
function of phase concentration results in unphysical sublattice populations
with this definition of the ideal mixing free energy density. Instead, we replace the
\ref{eq:pycalphad_ideal} terms with:
\begin{equation}
  y_i\text{plog}(y_i, \epsilon)
  \label{eq:moose_ideal}
\end{equation}
where \emph{plog} is the natural logarithm for $y_i > \epsilon$ with $\epsilon >
0$ and a Taylor expansion around $\epsilon$ for $y_i \le \epsilon$, as defined
in \cite{Schwen2017}. This substitution retains a strong driving force for
unphysical site fractions and pushes the solve back into the physical regime
while still avoiding the numerical issues associated with logarithms of negative
numbers. The choice of $\epsilon$ is a trade-off in which a low $\epsilon$
increases the stiffness of the equation system but more faithfully retains the
system's thermodynamic properties. Fig. \ref{fig:sublattice_concentrations}
(solid curves) shows the two sublattice concentrations for a successful
constrained minimization of the $\delta$-UZr free energy density. To obtain this curve,
we solve only Eqs. \ref{eq:phaseconc} and \ref{eq:sublatticechempot}, and
prescribe a linear concentration profile for the physical zirconium
concentration $c_{Zr}$.

\begin{figure}[tbp]
\centering
  \epsfig{width=\linewidth,file=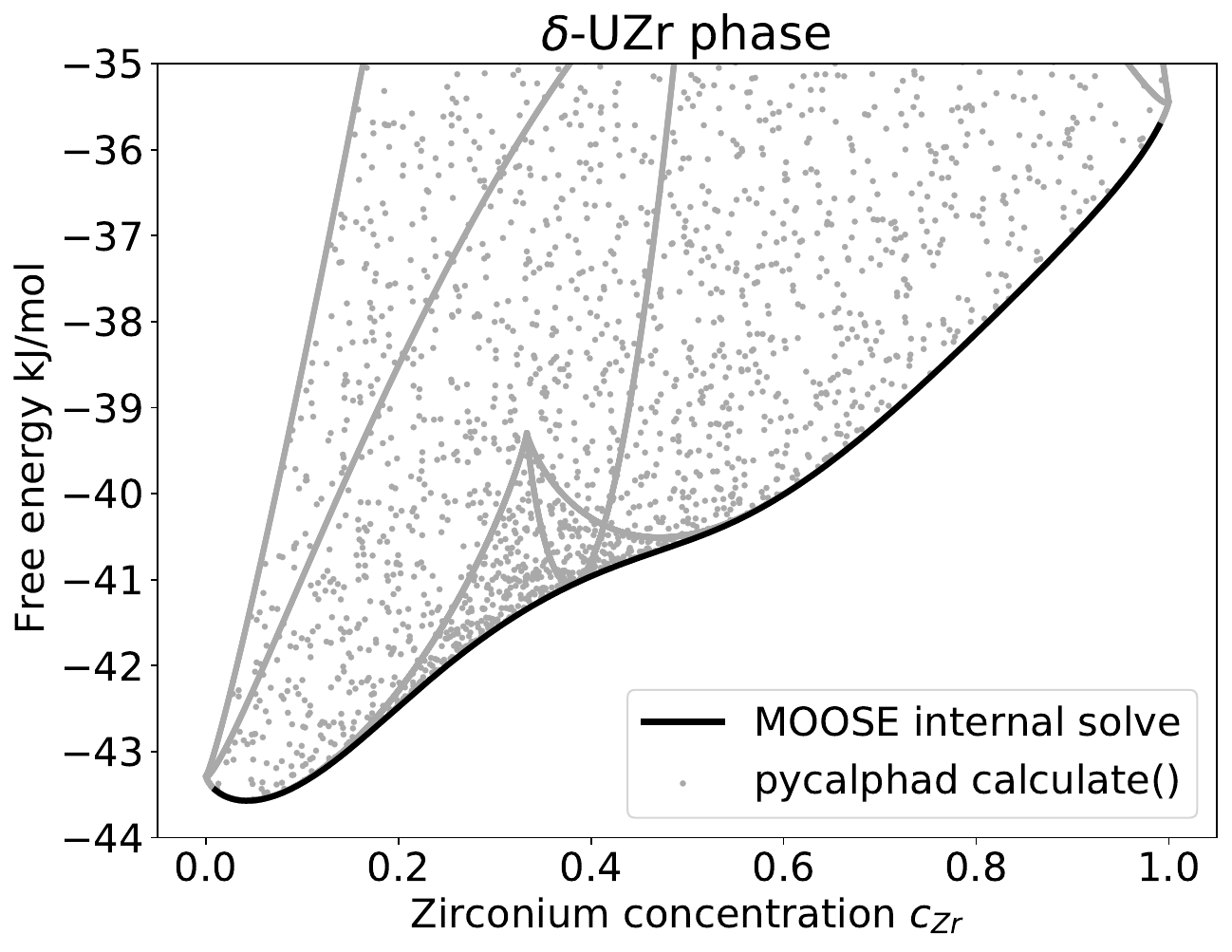}
  \caption{\label{fig:uzr_delta} Free energy vs. $c_{Zr}$ for varying sublattice concentrations. The data points generated by pycalphad result from random sublattice concentration values that satisfy a particular value of $c_{Zr}$. The data generated by MOOSE result from the constrained minimization of the phase free energy density for each value of $c_{Zr}$. As is seen from the plot, the MOOSE constrained minimization accurately captures the lower bound of the pycalphad free-energy data.}
\end{figure}

The resulting free energy density curve computed through the non-linear MOOSE solve is
plotted in Fig. \ref{fig:uzr_delta} (solid black line) on top of the pycalphad
scatter plot that samples the entire sublattice concentration space. The MOOSE
curve constitutes a lower bound (i.e., a constrained minimization of the the
phase free energy density).

To test the microstructural evolution of the UZr system within the SLKKS model,
we set up a sharp uranium (left) / zirconium (right) interface. For all phases,
we chose a gradient energy parameter $\kappa$ and barrier energy $W$ that
resulted in an interfacial width of $\Omega_U^\frac13$---where $\Omega_U$ is the
atomic volume of a uranium atom in the $\alpha$-uranium phase---along with an
interfacial free energy density of approximately 10 mJ/m$^2$. The selected interfacial
free energy density was low enough to allow for spontaneous formation of the
$\delta$-UZr phase. The atomic volumes of all three phases were set to
$\Omega_U$. We note that atomic volume changes during phase transformations can
be implemented through Eigenstrains, entailing a chemo-mechanical coupling. This
will be the subject of future work and was left out in this study in order to
focus on the chemical free-energy contribution enabled by the SLKKS model.

\begin{figure}[tbp]
\centering
  \epsfig{width=\linewidth,file=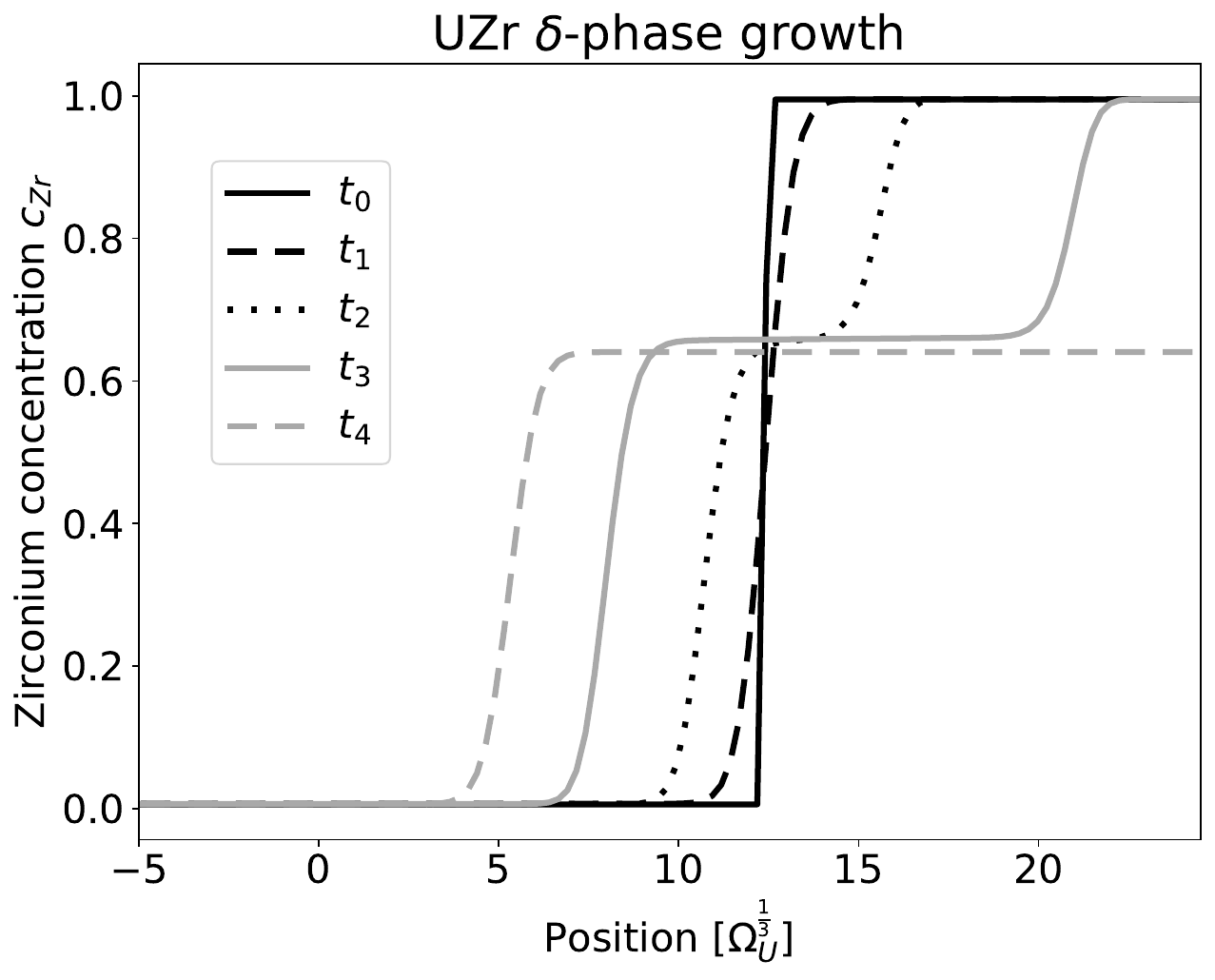}
  \caption{\label{fig:uzrphasefield} Evolution of the zirconium concentration at an initially ($t_0$) sharp uranium-zirconium interface. At time $t_1$, the interface profile evolves to the finite interface width of the phase-field model. At time $t_2$, a nucleus of the $\delta$-UZr phase at a zirconium concentration of $\approx0.65$ forms. The $\delta$-UZr phase grows ($t_3$) and fully consumes the $\alpha$-Zr phase on the right as the simulation reaches equilibrium at $t_4$.}
\end{figure}

Figure \ref{fig:uzrphasefield} shows the evolution of the interfacial profile
over time. The time units are arbitrary, and the diffusion coefficient $D$ was
set to unity. The interface starts off sharp at $t_0$. The non-conserved order
parameters are initialized with a sharp profile, as well. Within the first few
time steps, the interface softens ($t_1$), and the interfacial profile
determined by $\kappa$ and $W$ is established. At $t_2$, a nucleus of the
$\delta$-UZr phase at a zirconium concentration of $\approx0.65$ forms. We note
that, at higher interfacial free energies, the nucleation of this phase is
suppressed due to the energy barrier associated with the formation of an
additional interface. At $t_3$, the $\delta$-UZr phase has further grown,
consuming zirconium from the $\alpha$-zirconium phase on the right, which
vanishes at $t_4$.

\subsection{Molybdenum-nickel-rhenium}

The Mo-Ni-Re system assessed by Yaqoob and Crivello et al.
\cite{doi:10.1021/ic202479y,CRIVELLO2015233} was chosen to demonstrate a
ternary system with a particularly complex 5-sublattice model in the $\sigma$
phase with space group P42/mnm and a unit cell containing 30 atoms:
\begin{equation*}
  (Mo,Ni,Re)_2(Mo,Ni,Re)_4(Mo,Ni,Re)_8(Mo,Ni,Re)_8(Mo,Ni,Re)_8
\end{equation*}

\begin{figure}[tbp]
\centering
  \epsfig{width=\linewidth,file=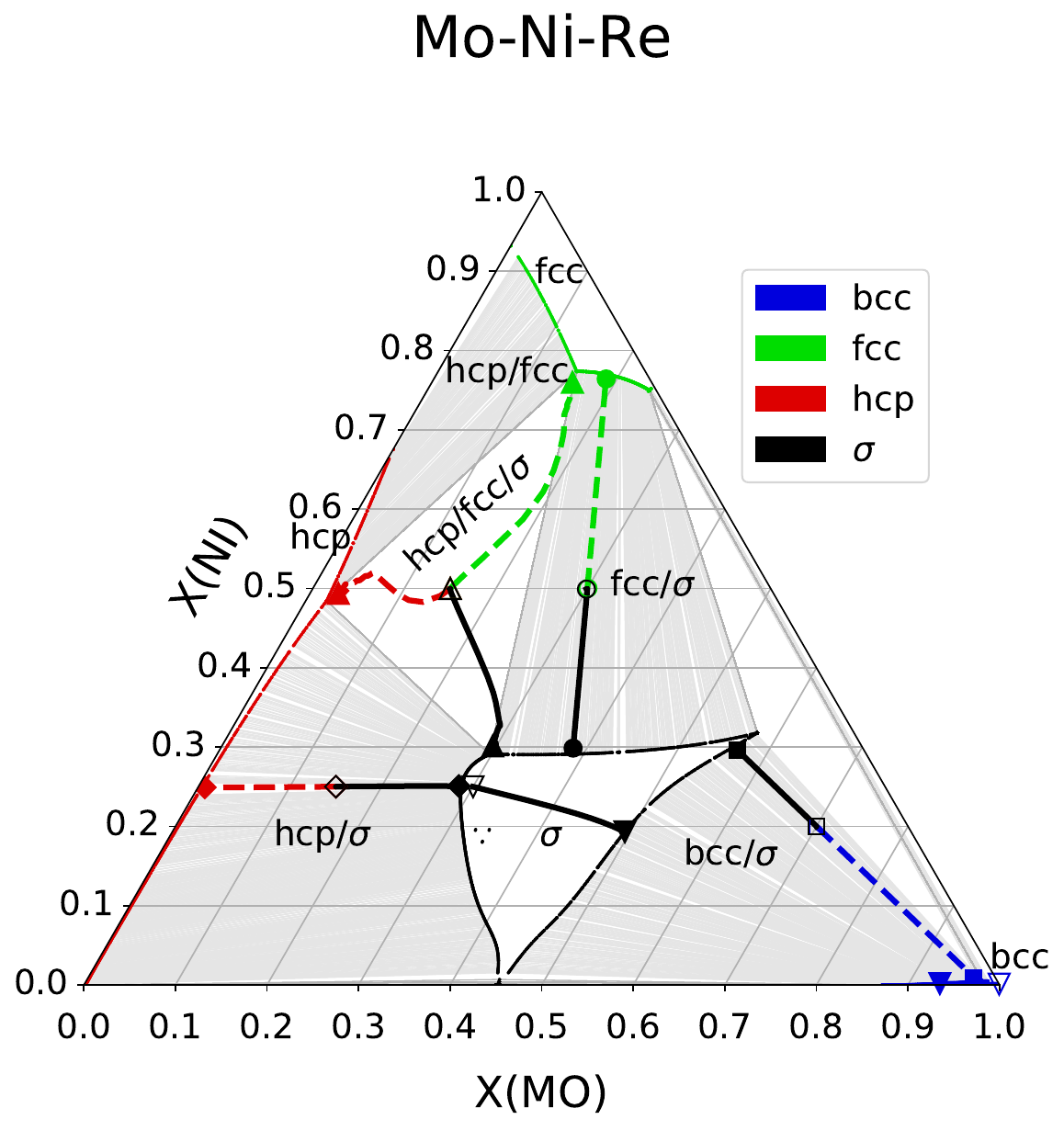}
  \caption{\label{fig:monirephasediag} Phase diagram of the ternary Mo-Ni-Re system. Note the Ni-rich fcc phase, the Mo-poor hcp phase, and the Mo-rich bcc phase. Located near the center of the phase diagram is the sigma phase with its complex sublattice structure. The superimposed solid and dashed trajectories show the compositional evolution of phases tracked in multiple phase-field simulations. Empty symbols denote starting compositions, and filled symbols denote compositions evolved to equilibrium.}
\end{figure}

We selected the temperature to be 900 K, at which the hcp phase is only stable
for molybdenum concentrations below about 3\%. This permits us to construct (for
demonstration purposes) two three-phase systems for which the thermodynamically
consistent switching functions have already been derived: one system containing
the fcc, bcc, and $\sigma$-phases, and one containing the fcc, hcp, and
$\sigma$-phases. We again set the interface energy to 10\,mJ/m$^2$, assigned the
same mobility to all components, and set the same molar volume for each phase.

Figure \ref{fig:monirephasediag} shows a phase diagram of the ternary Mo-Ni-Re
system at 900 K and under ambient pressure, as plotted using pycalphad. On the
phase diagram, we note the Ni-rich fcc phase, the Mo-poor hcp phase, and the
Mo-rich bcc phase. Located near the center of the phase diagram is the
$\sigma$-phase with its complex sublattice structure. Overlaid over the phase
diagram are the compositional trajectories of five phase-field simulations.

The initial compositions for each run are denoted by the empty symbols. The
filled symbols denote the final composition of each phase in the simulation. The
phase concentration is determined by computing the weighted average value of
each concentration variable, using the corresponding phase switching function
value as the weight. A switching function is one, in the region the
corresponding phase is active, and zero elsewhere. The interfacial regions can
introduce a small error that becomes negligible for thin interfaces and a small
interface-to-bulk ratio (i.e., a coarse microstructure). We note that, in this
multiphase model, the non-conserved order parameters do not directly correspond
to the phase fraction; instead, the above-mentioned switching functions---which
are functions of all the non-conserved order parameters---represent the phase
fractions.

\begin{figure}[tbp]
\centering
  \epsfig{width=0.7\linewidth,file=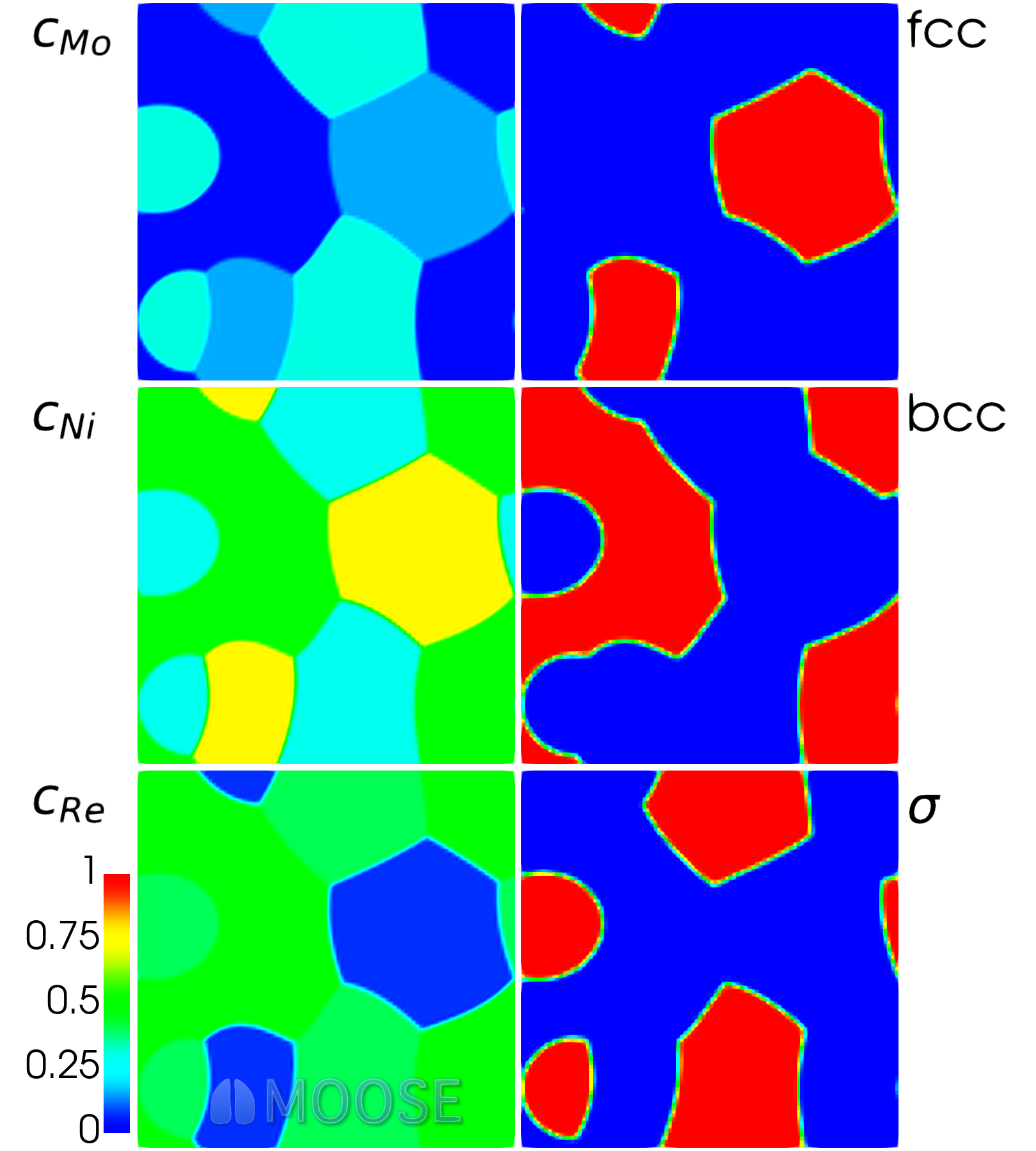}
  \caption{\label{fig:micro} Phase separated microstructure evolved from a homogeneous initial composition of Mo$_3$Ni$_{10}$Re$_7$, which is located in the hcp, fcc, and $\sigma$-phase coexistence region.}
\end{figure}

All simulations were run with an interfacial free energy density of 10\,mJ/m$^2$ and
equal mobilities for all components, as this work deals with the free energy density of
the bulk system. Interfacial effects and CALPHAD-informed mobilities will be the
topic of future work.

A diffusion couple with a $\sigma$-phase and a bcc phase side (downward-pointing
triangles) was set up with starting concentrations (open symbols) far inside the
respective phase regions, with Mo$_6$Ni$_5$Re$_9$ in the $\sigma$-phase and pure
molybdenum in the bcc phase. Through solute diffusion, the concentrations on
both sides approach the respective phase boundaries in excellent agreement with
the phase diagram. The underlying simulation was run in 1-D, thereby eliminating
interface curvature effects.

Three two-phase decomposition simulations were run, with concentrations in the
$\sigma$-fcc (circles), $\sigma$-bcc (squares), and $\sigma$-hcp (diamonds)
coexistence regions, as denoted by the gray tie lines in the phase diagram. Each
of the two-phase decomposition simulations was run in a finite-sized 2-D
domain. The converged simulations show good agreement with the phase diagram
boundaries. We note that the phase diagram does not take microstructural effects
such as interfacial tensions into consideration; thus, perfect agreement is not
expected.

A three-phase decomposition was run with a starting composition of
Mo$_3$Ni$_{10}$Re$_7$, located in the hcp, fcc, and $\sigma$-phase coexistence
region. An exemplary view of the ternary phase decomposition simulation
microstructure is shown in Fig. \ref{fig:micro}, with the Mo, Ni, and Re (top to
bottom) concentration order parameters on the left-hand side and the fcc, hcp,
and $\sigma$-phase fraction on the right-hand side. In a finite-sized domain, a
three-phase decomposition is not guaranteed to occur, as the energy penalty
resulting from the formation of additional interfaces associated with a third
phase can be prohibitively large.

\section{Conclusions}

As a natural extension of the KKS approach, we derived and demonstrated a
phase-field model that tracks sublattice compositions and allows direct use of
CALPHAD free energy models with multiple sublattices. The model, as presented,
permits an arbitrary number of constituents and sublattices. The model is easily
extendable to an arbitrary number of phases by utilizing the switching function
proposed by Moelans et al. \cite{MOELANS20111077} or Pogorelov and Kundin et al.
\cite{2013arXiv1304.6549P,KUNDIN2015448}, both of which suppress the formation
of spurious phases at interfaces.

\section*{Acknowledgements}
This work was supported through the INL Laboratory Directed Research \&
Development (LDRD) Program under DOE Idaho Operations Office Contract DE-AC07-
05ID14517. This manuscript was authored by Battelle Energy Alliance, LLC under
Contract No. DE-AC07-05ID14517 with the U.S. DOE. The U.S. Government retains
and the publisher, by accepting the article for publication, acknowledges that
the U.S. Government retains a nonexclusive, paid-up, irrevocable, worldwide
license to publish or reproduce the published form of this manuscript, or allow
others to do so, for U.S. Government purposes.

\section*{Data Availability}

The thermodynamic data used in the example cases for this study are available
online and can be accessed through TDBDB. The SLKKS model is implemented in the MOOSE\cite{moose-web-page}
phase field module\footnote{\url{https://github.com/idaholab/moose/blob/next/modules/phase_field/doc/content/modules/phase_field/MultiPhase/SLKKS.md}}.

\bibliography{ref}

\end{document}